\documentclass[aps,prb,twocolumn,showpacs,groupedaddress]{revtex4}
\usepackage[american]{babel}
\usepackage{color,graphicx,amsmath,amssymb,bm}

\begin{document}
\newcommand\ket[1]{\left|#1\right\rangle}
\newcommand\bra[1]{\left\langle#1\right|}
\newcommand\lavg{\left\langle}
\newcommand\ravg{\right\rangle}
\newcommand\be{\begin{equation}}
\newcommand\ee{\end{equation}}
\newcommand\matrixid{{\rm 1} \hspace{-1.1mm} {\rm I}}
\newcommand\stocavg[1]{{\lavg #1 \ravg}_{\textrm{stoc.}}}

\title{Full Counting Statistics of Cooper Pair Shuttling}
  
\author{Alessandro Romito}
\affiliation{NEST-INFM \& Scuola Normale
Superiore, Piazza dei Cavalieri 7, I-56126 Pisa, Italy}
\author{Yu.\ V.\ Nazarov}
\affiliation{Department of Nanoscience,
Delft University of Technology, 2628 CJ Delft, 
The Nederlands}
\date{\today}

\begin{abstract}
The Cooper pair shuttle is a simple model system that combines
features of coherent and incoherent transport.
We evaluate the full counting statistics (FCS) of charge transfer
via the shuttle in the incoherent regime.
We describe two limiting cases when the FCS allows for classical
interpretation. Generally, the classical interpretation
fails yielding negative and imaginary "probabilities".
This signals
that superconducting coherence survives
even in incoherent regime.
We evaluate the current noise in some detail.   
\end{abstract}

\maketitle
The Josephson effect~\cite{josephson62} consists in coherent 
Cooper pair transfer through a tunnel junction 
between two bulk superconductors.
It results in a
dissipationless current present as a ground state property 
of the system. This is in contrast to dissipative electron
transfer in normal conductors. In latter case, it is possible
to evaluate the full counting statistics (FCS) of charge transfers
~\cite{levitov96} in terms of classical probabilities.
Thereby, the transport properties are fully understood and 
characterized: one can predict
not only the average current, but also the current noise and
all higher moments of the current distribution 
function~\cite{nazarov_book02}.      

One can also access the FCS
of a superconducting Josephson junction~\cite{belzig01}.
Due to gauge symmetry breaking in the superconducting state, 
this FCS 
can not be interpreted in classical terms yielding
negative "probabilities".
The use and the interpretation 
of the FCS in this case is that it determines
the quantum evolution
of the detector that measures the current~\cite{kindermann02}. 

A recent proposal puts forward an interesting way to transfer 
Cooper pairs: to shuttle them (controllably)
between the superconducting 
electrodes~\cite{gorelik01}. 
Although the original proposal
puts emphasis on mechanical degrees of freedom,
a Cooper pair shuttle is essentially  
a superconducting single electron transistor (SSET) with variable 
time-dependent Josephson coupling to the superconducting
leads~\cite{romito03}.
The Coulomb island of this SSET is 
brought in contact with one electrode at a time, the electrode contacted
being  periodically altered~\cite{shekhter03}. 

It is interesting that the supercurrent between 
the leads is achieved as a result of a non-equilibrium driven 
process, generally accompanied by dissipation. 
The coherence of Cooper pairs transferred is determined 
by the coherence of different charge states in the island,
this being mostly affected by fluctuations of the gate voltage~\cite{romito03}. 
The transport results from the interplay between the 
coherent Josephson coupling and decoherence effects.
This is why the Cooper pair shuttle
presents a model to bridge
between the limits of coherent and incoherent
transport. The model is simple indeed, one can 
restrict the consideration to just two quantum states.

Thus motivated, we have analyzed the FCS of
the Cooper pair shuttle
in the most interesting 
{\it incoherent} regime, where no net supercurrent is shuttled. 
The regime is achieved in the limit when the 
voltage fluctuations are {\it classical}.  
If the fluctuations come from an environment, this implies 
that the temperature
of the environment exceeds the relevant
energy scales of the shuttle.
Naively, from the absence of the net current 
one would conclude that no charge transfer occurs in the system. 
However, transfers do occur,
 the current is zero only in average,  
and the FCS presents a convenient way to reveal this circumstance.
This stipulates the understanding of transport properties of the
shuttle and, generally, the interplay between the coherent and
incoherent transport.

The results are as follows. 
If the period of the shuttling is sufficiently long for decoherence
to be accomplished, the FCS can be interpreted in terms of 
classical elementary events: Cooper pair transfers.
During the shuttling cycle, either no transfer takes place or
a pair is transferred in either direction.
There is an apparent similarity with the FCS of
the pumping in normal systems
studied in~\cite{levitov01,andreev00}.
The FCS in the opposite limit of short cycles
allows for alternative classical interpretation
in terms of 
a superconducting current randomly
switching between two opposite values.

In the general intermediate situation,
the FCS can not be interpreted in classical
terms. An attempt to evaluate the probabilities per cycle yields 
negative and even imaginary values. This is a clear signature
of the fact that the superconducting coherence survives strong dephasing
although this coherence does not manifest itself in net superconducting 
current. It was recently explained~\cite{tobiska03} that any FCS can be 
characterized directly so one would not have to measure
higher cumulants of the current noise one by one. However, 
the immediate physical value measured would also depend on the properties 
of the concrete detector. This is also a way 
to experimentally observe the FCS of the Cooper pair shuttle.

\begin{figure}
\begin{center}
\includegraphics[width=2.5in]{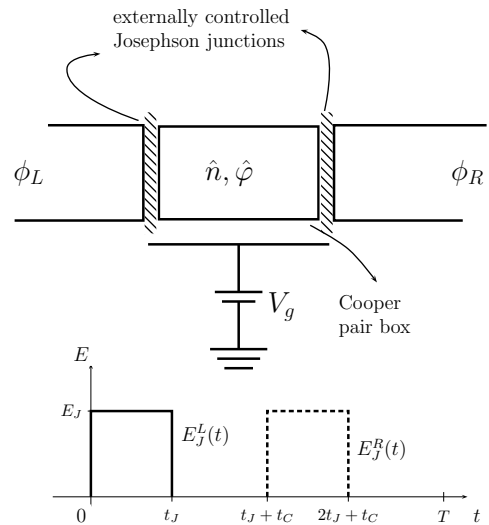}
\end{center}
\caption{Upper panel.
Cooper pair shuttle consists of a Cooper pair box
coupled to superconducting
leads through Josephson junctions.
Two charge states are tuned to degeneracy point by
the gate voltage. Lower panel.
The specific time dependence of $E^{(R,L)}_J$
within a single period $T$ provides the shuttling.}
\label{system}
\end{figure}

The system of interest is 
presented schematically  
in Fig.~\ref{system}. It consists of a Cooper pair box,
or Coulomb island,
connected with Josephson junctions to the
superconducting leads $L$ and $R$.
Conventionally,~\cite{gorelik01,romito03,shekhter03}
 we assume the separation of energy scales,
$\Delta \gg E_C \gg E_J$, $\Delta$ being the superconducting energy
gap, $E_C$ being the charging energy of the box, and $E_J$ being a typical
Josephson junction energy. Under these conditions 
quasiparticles degrees of freedom are not involved in the system's dynamics. 
In addition, the gate voltage
is chosen to bring two charge states of the box, say,
 $|0>$ and $|1>$ to the degeneracy point.
Under these assumption, we can restrict the consideration
to these two states only and the shuttle is described by the Hamiltonian 
\be
\hat H(t) =
2e V(t) \sigma_z - \sum_{b =L,R} \frac{E_J^{(b)} (t)}{2}
\left( e^{i\phi_b} \sigma_+ + \sigma_- e^{-i\phi_b}\right) \, ,
\ee 
which we write in terms of $2 \times 2$ Pauli matrices $\sigma_z$,
$\sigma_{\pm} \equiv (\sigma_x \pm \sigma_y)/2$
in the space spanned by $|0>,|1>$.
$V(t)$ is proportional to the deviation of the gate
voltage from the value corresponding to the degeneracy point and
$\phi_R$, $\phi_L$ are the phases of the macroscopic 
superconductors. Only $\phi=\phi_R-\phi_L$ is a physical quantity
which we will assume to be fixed; this can be obtained by 
closing the circuit endpoints in a loop pierced by a constant 
magnetic flux.
$E_J^{(L,R)}(t)$
are time-dependent Josephson energies of left and right 
junction.
For the sake of concreteness, 
we assume stepwise periodic variation
of both Josephson energies with
time as shown in the lower panel of Fig. 1.
Each lead is contacted during time interval $t_J$ and the box contacts
no lead during time interval $t_C$ for each period $T=2(t_J+t_C)$.
We call $t_J$ and $t_C$  the ``Josephson contact'' 
and the ``free evolution'' time respectively. 
An idea to realize the required time 
dependence in Josephson coupling by means of an
experimentally realizable SQUIDs based device has been 
proposed in Ref.~\onlinecite{romito03}.  

We assume $V(t)$ to be a classical stochastic variable 
with white noise statistics,  
${\lavg  V(t) \ravg}_{\textrm{stoc.}} =V_0$ and 
${\lavg V(t) V(t') \ravg}_{\textrm{stoc.}} = \gamma \hbar^2/(2e)^2 \delta(t-t')$,
where $\stocavg{\cdot}$ defines the average over 
the fluctuations. Thus defined, $\gamma$ is just the inverse
decoherence time of the two charge states.
If we neglected the fluctuations, the time evolution of the system
would be fully coherent governed by the coherent
part of the Hamiltonian $\hat H_c \equiv {\lavg\hat
H\ravg}_{\textrm{stoc.}}$.
In this case the quantum state of 
the central grain would 
acquire dynamical phases $2 \theta= E_J t_J/\hbar$ during the Josephson contact 
times and $2 \chi= 2eV_0t_C/\hbar$ during the free evolution
times (we assume that $V_0$ is only present during
both intervals of the free evolution time).

With fluctuations, the shuttle will be described
by a $2 \times 2$ density
matrix that obeys the following Bloch equation:
\begin{equation}
\frac{\partial{\hat\rho}}{\partial t}=
-\frac{\imath}{\hbar}
 \left( \hat H_c(t) \hat\rho - \hat\rho\hat H_c(t)\right) 
-2 \gamma \left( \hat\rho - \sigma_z \hat\rho \sigma _z \right) \, .
\label{Bloch}
\end{equation}

The only stationary solution of this equation is quite trivial: 
$\hat\rho  \propto \hat 1$,
this corresponds to the absence of the averaged superconducting current.
Therefore, the shuttle operates in incoherent regime. This
is a combined effect of the decoherence term and Josephson coupling.
In the absence of Josephson coupling, voltage 
fluctuations can not cause transitions between the charge states 
so no relaxation takes place. With Josephson coupling, the voltage 
fluctuations cause transitions between the stationary states separated
by energy $2E_J$ at $V_0=0$. Classical voltage fluctuations
result in equal transition rates with increasing and decreasing
energy. In fact, the ratio of these rates is given by Bolzman factor
$\exp(2E_J/T_b)$, $T_b$ being the temperature of the environment
producing the fluctuations. As shown in Ref.~\onlinecite{romito03}, proper
account of this factor leads to anisotropic $\hat\rho \ne \hat 1$
and to the net supercurrent vanishing at $T_b \gg E_J$.
This defines the incoherent regime under consideration. 
  
Let us evaluate the FCS. 
One starts with the assumption 
that a transport process  can be characterized by the
probabilities $P_{\tau}(N)$ of $N$ electrons transferred through the
contact in a 
time interval $\tau$ and attempts to compute the characteristic function of this
probability distribution defined as~\cite{levitov96}
\be
e^{-\mathfrak{S}(\lambda,\tau)}= \sum_{N} P_{\tau}(N) 
e^{\imath \lambda N}\, .
\label{old_fcs}
\ee
The general quantum expression for this function can be obtained
by coupling the system to a detector for a measuring time $\tau$ and interpreting 
the detector readout in terms of charge transfer~\cite{kindermann02}.
The FCS in this case is defined as integral kernel that relates
the initial and final density matrices of the detector, and reads
\be 
e^{-\mathfrak{S}(\lambda, \tau)} =
\textrm{Tr} \underbrace{\left[\mathcal{U}_{+\lambda}(\tau,0) \, \hat\rho(0) \,
\mathcal{U}^{\dag}_{-\lambda}(\tau,0) \right] }_{\hat\rho '(\tau)} \, , 
\label{fcs_general}
\ee
where $\mathcal{U}_{\lambda}(\tau,0)=\overrightarrow{T}e^{-\imath \int_0^{\tau} \, 
\hat H_{\lambda}(s)\, ds }$ is the unitary 
evolution operator  corresponding to 
the modified Hamiltonian 
$\hat H_{\lambda} = \hat H + \lambda \hat I/e$, $\hat I$ being the operator
of the electric current. Two such operators provide non-unitary evolution
of the initial density matrix of the system $\hat\rho(0)$
into $\hat\rho'(\tau)$. 

For our shuttle model, the implementation of this general scheme is
especially simple since the density matrix is just $2 \times 2$ matrix.
Following Ref.~\onlinecite{belzig01}, we also gauge the counting field $\lambda$ to 
the left electrode, this yields 
$\hat H_{\lambda} = \hat H_c(\phi_L \rightarrow \phi_L + \lambda)$. We derive an equation for 
$\hat\rho'$ that looks very similar to the Bloch equation (\ref{Bloch})
\begin{equation}
\frac{\partial\hat{\rho'}}{\partial t}=
-\frac{\imath}{\hbar} \left( \hat H_{+\lambda} \hat\rho' - \hat\rho'
\hat H_{-\lambda}\right) 
-2 \gamma \left( \hat\rho'- \sigma_z \hat\rho'\sigma _z \right) \, .
\label{Bloch2}
\end{equation}
The difference is that the Hamiltonians governing
evolution of "bra"'s and "ket"'s differ
by opposite $\lambda$ shifts. 

This  system of 4 linear equations can be solved
with initial conditions $\hat\rho'(t_0)$ 
to obtain a linear map $M$ that gives $\hat\rho'$
after a cycle, 
$\hat\rho'(t_0+T)=M \hat\rho'(t_0)$ 
All four eigenvalues $\mu_i$ of $M$ satisfy the condition
$|\mu_i| \in [0,1]$. 
Since we study statistics of low-frequency fluctuations, 
 $\tau \gg T$ by definition. 
Therefore, the FCS is determined by the eigenvalue 
with the greatest 
magnitude,
\begin{equation}
e^{-\mathfrak{S}(\lambda, \tau)} = \bar{\mu}(\lambda)^{\tau/T} \, ,\hspace{0.2cm} 
\textrm{where} \hspace{0.2cm} 1-\bar{\mu} = \min_{i=1,\dots,4}(1- \mu_i) \, .
\label{final_fcs}
\end{equation}
This is in accordance with the method and the result~\cite{bagrets02}
for FCS of charge transport described by
a master equation.

The most transparent way to present the FCS is to define the
probabilities $p_N$ to transfer N electrons {\it per cycle}, as it has been done
in Ref.~\onlinecite{levitov01} to characterize the pumping
of normal electrons,
\begin{equation}
p_N = \frac{1}{2 \pi} \int_{-\pi}^{\pi} d \lambda e^{-i N\lambda} \bar{\mu}(\lambda) \, .
\end{equation}
The FCS in the limit considered is always an even  
function of $\lambda$, so that the pairs are 
shuttled in either direction
with equal probability, $p_{N} = p_{-N}$.

We analyze the FCS  in several limiting cases.
First we consider the limit of long periods
$1/t_C, 1/ t_J \ll \gamma \ll E_J/\hbar$.
This is in fact an adiabatic limit since the relaxation
and decoherence are fully developed within each time
interval $t_C, t_J$. In this case, 
\begin{equation}
\bar{\mu} (\lambda) = \cos^2 \lambda
\; \rightarrow \; p_0=\frac{1}{2}; p_{\pm 2}=\frac{1}{4} \, ,
\label{fcs_strong}
\end{equation}
so that, each shuttling between the superconductors
transfers either one Cooper pair or none, this occurs
with equal probabilities.
The pair is transferred with equal probabilities
in either direction.
It is interesting to note that this simple result is quite
general and  relays on neither the periodicity
of shuttling nor the concrete time dependence
of $E_J(t)$ provided the adiabaticity is preserved.
The shuttle may even return (several times)
to the same superconducting
terminal before contacting the opposite one: the charge
transfer is in this case associated with two trips
between the opposite terminals.
Leading corrections to adiabatic FCS are exponentially small,
$\simeq \exp(-2t_J\gamma)$.

Another interesting limit is that of 
small Josephson
couplings $E_J \ll \hbar \gamma$ and sufficiently long cycles
$\hbar/t_C,\hbar/t_J \ll E_J $. 
In this case, the charge relaxation
time $\simeq \gamma\hbar^2/ E_J^2$ may be long,
exceeding both decoherence time and cycle duration.
The relevant parameter $\exp(-t_J E_J^2/(\hbar^2 \gamma)) \equiv f$
determines the efficiency of the charge relaxation
during a cycle.
The FCS becomes more complicated, for $f \simeq 1$ it is given by
\begin{eqnarray}
& & \bar{\mu} (\lambda) 
=\frac{1}{2}  (1+f^2)\cos^2\lambda + f\sin^2\lambda +
(1-f^2) \times \nonumber \\
& & \times \frac{1}{2}\sqrt{\cos^2\lambda \left( 1 - \left( \frac{1-f}{1+f}\right)^2 
\sin^2\lambda \right)} \, . 
\label{fcs_low}
\end{eqnarray}
and, generally speaking, all $p_N \ne 0$. This is because
the charge can not completely relax during a cycle providing 
"memory effect" so that charge transfers 
in different cycles are not independent  
and elementary event of
charge transfer can encompass several cycles.
Still, probabilities remain positively defined owing to the fact 
that in
the  limit of long cycles the FCS depends neither on the
superconducting phase $\phi$ nor on 
dynamical phases $\theta, \chi$. 

\begin{figure}
\includegraphics[width=2.5in]{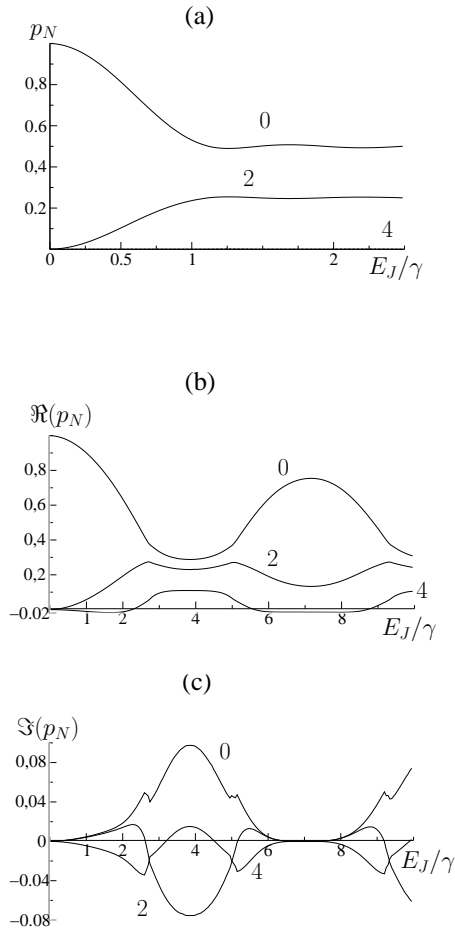}
\caption{
Fourier components, $p_N$, with $N=0,2,4$ 
of $\bar{\mu}(\lambda)$
plotted versus $E_J/\hbar \gamma$. Each line is labeled by 
the corresponding value of $N$. 
Adiabatic regime (a): $\gamma t_J = 5 $; 
positive probabilities. 
General situation (b-c): $\gamma t_J =0.9$;
$p_N$
can be negative and even immaginary.
Other parameters are fixed to 
$t_J=t_C$, $\chi=\pi /5$, $\phi=\pi/4$ for all plots.
}
\label{Pplot}
\end{figure}

Beyond this limit, the FCS  does depend on $\phi$.
As mentioned in Ref.~\onlinecite{belzig01}, the classical
interpretation in this case may fail 
and one can not assure that $p_n$ are positive or
even real.
Indeed, we have found this in numerical calculations
at $\gamma T \simeq 1$ (Fig. \ref{Pplot}).
This clearly signals that the superconducting
coherence remain in the system although the decoherence
completely destroys the average supercurrent.

This becomes evident when analyzing the opposite
limiting case of very short cycles $\gamma T \ll 1$.
In this case, the cumulants of the current are
contributed by FCS at $\lambda \simeq \gamma T \ll 1$.
In this region, the FCS reads
\be
\frac{\bar \mu -1}{T} \simeq -\frac{\tilde{\gamma}}{2}
+\sqrt{
\left(\frac{\tilde{\gamma}}{2}\right)^2
-\left(\frac{\lambda I_s}{e}\right)^2
} \, ,
\ee
where $I_s$ is the superconducting current in the absence
of decoherence $\gamma =0$ and $\tilde\gamma$ is the
relaxation time in the limit $\gamma \rightarrow 0$:

\begin{equation*}
I_s = \frac{2e}{T} \frac{\partial \alpha}{\partial
\phi}  ,\hspace{.5cm} 
\tilde \gamma = \frac{\gamma}{\sin^2 \alpha}
\left( \frac{t_J}{T}P_J+ 
\frac{t_C}{T}P_C \right) 
\end{equation*}
where  $\alpha \equiv\arccos \left( \cos^2\theta \cos2\chi -\sin^2\theta\cos\phi\right)$ 
and  $P_J$, $P_C$ are always positive polynomial functions in $\sin(\cdot)$ and $\cos(\cdot)$ 
of the phases $\phi, \chi, \theta$:
\begin{eqnarray*}
P_C(\phi,\chi,\theta) &=& 2 \sin^2(2\theta) \left( 1+\cos\phi \cos(2\chi) \right) \\
P_J(\phi,\chi,\theta) &=& 2 \left( 1+ \sin^2 \theta \cos\phi \cos(2\chi)\right) +\\
 & -& \cos(2\theta) \left( \cos^2 \theta \cos^2(2\chi)- \sin^2\theta \cos^2\phi \right)
\end{eqnarray*}
FCS of this type corresponds~\cite{belzig01} to a
supercurrent randomly switching at time scale
$1/\tilde \gamma \gg T$ between the opposite
values $\pm I_s$, so that the coherence is preserved
at this time scale. The charges are transferred in long trains
of $\simeq 1/\tilde \gamma T$ of elementary charge,
this leads to significant low-frequency current noise.

\begin{figure}[ht]
\includegraphics[width=2.5in]{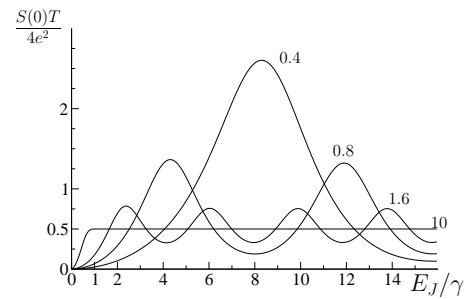}
\caption{Low-frequency noise 
in the intermediate regime for different values of 
$\gamma t_J$, as expressed by each line label. 
The noise and its oscillations with
$\theta$ increase with decreasing $\gamma t_J$.
Other parameters are fixed to 
$t_J=t_C$, $\chi=\pi /5$, $\phi=\pi/3$ for all plots.
}
\label{noiseplot}
\end{figure}

Low-frequency noise $S(\omega=0)$ is obtained
directly from the definition of the FCS,
$S(0) = - e^2/T \lim_{\lambda \rightarrow 0}
\partial^2( \bar \mu /\partial \lambda^2)$.
From the Eq. (\ref{fcs_low}) we obtain
$S(0)=2e^2 /T \times (1-f)/(1+f)$ in the adiabatic limit. 
The noise is enhanced in the opposite limit of short cycles,
$S(0) = 2 I^2_s/\tilde \gamma \gg e^2/T$ and is sensitive
to all dynamical phases.
The numerical results in the intermediate regime (Fig.
\ref{noiseplot}) show in addition a quasi-oscillatory dependence on
the dynamical phase $\theta$. 

The authors appreciate many highlighting
discussions with R. Fazio. This work was supported by the 
EU (IST-SQUBIT, Grant No. HPRN-CT-2002-00144) and by Fondazione
Silvio Tronchetti Provera.


\end{document}